# A Framework for Performance Evaluation of ASIPs in Network-based IDS

Majid Nezakatolhoseini[1] and Mohammad Amin Taherkhani[2]

[1]Department of Computer and Mechatronic, Science and Research Branch Islamic Azad University, Tehran, Iran
`m.nezakat@srbiau.ac.ir`
[2]Department of Computer Engineering, Shahid Beheshti University, Tehran, Iran
`m_taherkhani@sbu.ac.ir`

## Abstract

*Nowadays efficient usage of high-tech security tools and appliances is considered as an important criterion for security improvement of computer networks. Based on this assumption, Intrusion Detection and Prevention Systems (IDPS) have key role for applying the defense in depth strategy. In this situation, by increasing network bandwidth in addition to increasing number of threats, Network-based IDPSes have been faced with performance challenge for processing of huge traffic in the networks. A general solution for this bottleneck is exploitation of efficient hardware architectures for performance improvement of IDPS. In this paper a framework for analysis and performance evaluation of application specific instruction set processors is presented for usage in application of attack detection in Network-based Intrusion Detection Systems(NIDS). By running this framework as a security application on V850, OR1K, MIPS32, ARM7TDMI and PowerPC32 microprocessors, their performance has been evaluated and analyzed. For performance improvement, the compiler optimization levels are employed and at the end; base on O2 optimization level a new combination of optimization flags is presented. The experiments show that the framework results 18.10% performance improvements for pattern matching on ARM7TDMI microprocessors.*

## Keywords

*Intrusion Detection, Hardware Architecture, Network based IDS, ASIP, Optimization in Compiler level*

## 1. Introduction

More than three decades has been passed from introducing of network surveillance concept as a basic idea for intrusion detection systems by James Anderson [1] in 1980. After this start point, Denning [2] presented the first model of IDS and caused the researches on intrusion detection were accelerated in next years. The evolution of intrusion detection systems is currently in a state which this kind of security tool has become as an integral part of information technology infrastructures.

In one hand, network infrastructures have been growing in recent years and this causes increasing the network traffic. On the other hand, reports from Computer Security Incident Response Teams (CSIRT) show that the number of threats and also complexity of attacks have been dramatically increased [3]. Based on these conditions, NIDS have faced with the challenge of real-time processing of huge network traffic. A general solution for overcoming to this challenge is exploitation of hardware accelerated architectures. Deferent kinds of hardware platforms including General Purpose Processors (GPP), FPGA and Application Specific Instruction-set Processors (ASIP) and Application Specific Integrated Circuits (ASIC) have been used for improvement of NIDS performance.

In this research, a framework is proposed containing expandable and efficient microprocessors for implementation of NIDS. Development of this framework is due to: 1) flexibility in system reconfiguration and 2) performance improvement. Note that the networks are vulnerable to new attack patterns, so updating the attack patterns in NIDS is inevitable. According to this limitation, it is very dramatic for many hardware platforms to compatible with this requirement. Based on this challenge we attempt to provide a framework for implementation of performance-critical section of intrusion detection engine on ASIPs as a flexible, reusable, high-performance and low cost platform.

The contribution of this paper is listed as follow:

- A taxonomy of hardware architectures for network-based IDS is presented.
- A framework for performance evaluation of popular microprocessors is proposed.
- A new combination of compiler optimization flags has been introduced.

The paper is organized as follow: In section 2, related researches have been reviewed and a taxonomy of hardware architecture for NIDS is introduced. The proposed framework is presented in section 3. Section 4 describes the optimization levels in compiler. In section 5, Implementation and experimental results are shown and finally section 6 concludes this paper.

## 2. TAXONOMY OF RELATED WORKS

In this section, the researches which focused on high-performance IDPS, have been classified. Based on the classification, a taxonomy of hardware architecture for NIDS has been presented. The taxonomy is shown in Figure 1. As shown in the Figure 1. the hardware architectures can be reviewed based on two deferent viewpoints: presented hardware platforms and proposed structures.

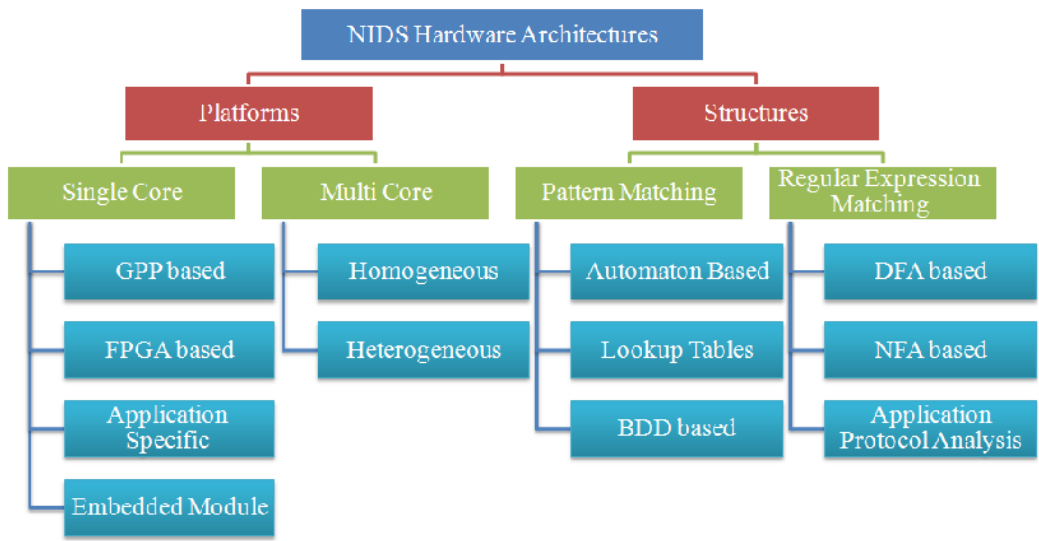

Figure 1. A Taxonomy of hardware architectures for high performance NIDS

### 2.1. Hardware Platforms

The taxonomy of hardware architectures is started by classification of pervious researches in viewpoint of hardware platforms.

As shown in the Figure 1. , NIDS hardware platforms could be divided in two subcategories based on number of processing elements required to perform the overall or specific functionality of NIDS. Considering the future architecture could be mapped to multiple cores, the pervious architectures are classified in the single processing core. In [4] Paxson et. al. a multi-core architecture is proposed for concurrent execution of network event processing. Also, in [5] parallel design has been presented by using heterogeneous platform based on multi-core CPU in addition to Graphical Processing Units (GPU).

Single Core platforms could be classified in the following sub-categories:

### 2.1.1. General Purpose Processor

In general, all software based implementation of network intrusion detection systems are classified in this category. In this group, there are some NIDSs which run in user land. Many open source network intrusion detection systems such as Snort [6] and Bro [7] are in the user space and as a result are classified in this category.

In behavioural viewpoint, many classic algorithms and general techniques for pattern matching and regular expression matching in NIDSs are used to implement on a general purpose processor. Primitive versions of snort used Booyer-Moore algorithm [8] and the newer version uses an algorithm based on aho-corasick [9] approach.

### 2.1.2. FPGA

Performance requirement and need for reconfigurable platforms (due to NIDS rule changes) cause many researches focused on FPGA as a platform for their IDS designs. The proposed structures in [10,11,12] are implemented on FPGA platform. In addition, the intrusion detection system introduced by Katashita et al. [13] is implemented on FPGA. Although some researches such as [10] discussed their structure could be implementable on the other platforms such as general purpose processors or ASIC, but their result and evaluation is based on FPGA platforms. Another interesting feature of FPGA platforms is reconfigurable capability which motivates the researches to implement their solution on this platform.

### 2.1.3. Application Specific

In addition to general purpose processors and FPGA platforms, we could name a platform, which is used for specific application such as network inspection and intrusion detection. In a top level of this category, there are application specific instruction-set processors (ASIPs). In [14] an architecture is proposed with the name of Keyword Match Processor. However, a FPGA is used for simulation and initial implementation the architecture. On the other side of platforms in the application specific spectrum, there are Application Specific Integrated Circuits (ASICs) which are designed to operate IDS functionality. The most important advantage of this platform is performance satisfaction and low power properties. Similar to multi cores platforms, there is a critical gap in pervious researches which capable to analysis and synthesis an efficient architecture on application specific platforms.

Recently benefits of exploitation of processing power in Graphical Processing Unit (GPU) are considered for performance critical applications in network security, especially network based intrusion detection and prevention systems [15,16]. This kind of platform can also be considered in the application specific category.

### 2.1.4. Embedded Modules

In addition to above platforms, in some related works different hardware modules are proposed to implement a specific component of NIDS. Memory is one of interesting embedded modules

International Journal of Network Security & Its Applications (IJNSA), Vol.4, No.5, September 2012

in those works. Bloom filters in [17] are example of the structures which mapped to internal and external memories to improve pattern matching effectiveness. Also the proposed structure in [18] is implemented on static memories.

In addition to general purpose memories, Content Addressable Memories (CAM) became an important hardware module to improve the IDS performance. CAM modules had been used in other network services such as IP forwarding and packet classification in routers and other network devices [19]. An advantage of CAM based platforms is parallel search capability [20]. Therefore, in some solutions for IDS pattern matching is used CAM structures to improve the required performance. Although CAM platform have some problems to handle arbitrary pattern matching, but some researches have attempted to propose techniques to solve the problems. CAM based architectures are classified in the following categories:

- **Binary CAM:** Each cell in a binary CAM could be 0 or 1. In comparison to Ternary CAM this module have less flexible and there are rare works such as [21] which uses this platform to handle pattern matching in intrusion detection systems [22].

- **Ternary CAM:** Each cell in a ternary cam could be in 0, 1 or don't case (x) state. The latter state is useful for matching of case insensitive patterns. In [22] this technique is used to improve the speed of matching in IDS and Antivirus services. In [23] a TCAM based platform is presented which capable of matching multi state regular expressions in minimum records of Ternary CAM.

## 2.2. Structures

The hardware architectures for acceleration of application of deep packet inspection in intrusion detection systems could also be categorized from view point of structure. As shown in figure 1, the structures are proposed because of two kinds of concerns: Pattern Matching and Regular Expression Matching.

### 2.2.1. Structures for Pattern Matching

In general, the pattern matching problem is describes as follow: to finding a non-null set of patterns $\{p_0, p_1, …, p_n\}$ in an input sequence *S*. Whenever any of patterns is found in the input sequence, a match is occurred. The extended version of pattern matching (for matching multiple patterns with offset) is used in engine of many network based IDS. This type of pattern matching is a time-consuming component. As a result, many researches have been studied on acceleration of pattern matching as an important component of Network based intrusion detection systems [23].

Generally, related researches on pattern matching can be categorized in the following structures:

- **Automaton based structures**: The main concern in the following researches is the pattern matching and their structures are based on state automata. Some of these are as follows: In [24], a GB pattern matching tool that supports fully TCP/IP network has been described. This system divides TCP/IP stream to sub-streams and distributes the load to several pattern matching units which use Deterministic Finite state machine (DFSM) pattern matching. When the number of rules increases, the number of states required to implement methods based on DFA is significantly increased. This can lead to reduce the performance of these systems. Tan et al. in [10] proposed state automata known as Bit-split finite state machine to improve the performance of intrusion detection systems. In addition, TDP-DFA [25], CDFA [26] can be categorized in automaton based structures.

- **Lookup tables based**: Lookup table based structures was introduced to improve performance of pattern matching techniques in intrusion detection process. The main idea behind usage of these structures is reservation of patterns in table with specific format and in the next step parallel search is executed. The research which used lookup tables structures, is classified in two subcategories:

- o Raw Table Structure
- o Hash Table Structure

In the first subcategory, generally the unrefined pattern is stored into tables. Although in some cases the pattern may be split to fit in table rows, but the content of patterns not to be changed. The work discussed in [22] is an instance of researches classified in raw table structure category.

In the second subcategory, content of patterns are refined and modified using specific hash functions. Then the modified patterns are stored in reference table. Bloom filters are categorized to this kind of structure. The Bloom filters [27] as an efficient estimate memory have been used in the field of research related to pattern matching in intrusion detection. Bloom filters use a random technique for testing membership queries in the set of strings. Predefined set of signatures that have been grouped according to their length are stored in set of parallel Bloom filters in the form of hardware. Each of these Bloom filters includes the special length signatures.

- **BDD based**: The third class in pattern matching solutions describes Binary Decision Diagram (BDD) based structures. In [28] a pattern matching engine is proposed based on BDD structures. In their methodology, raw patterns extracted from attack signatures, in the next phase the extracted patterns are converted to Boolean expressions. The next phase in their methodology discusses transformation of BDD to ROBDD. Then, they have attempted to partition and optimized the compound BDD. The resulted structure is implemented into content addressable memories.

### 2.2.2. Structures for Regular Expression Matching

Recently, increasing complexity of attacks against network resources causes intrusion detection systems have to use regular expressions (RE) for describing the attack signatures. This change caused RE matching to become a critical bottleneck for engine of NIDS. Usually, Deterministic Finite state Automaton (DFA) and Nondeterministic Finite state Automaton (NFA) are used for analysis of regular expressions. Traditional or uncompressed DFA problems which are mentioned above lead to other DFA including compressed DFA, $D^2FA$ [29], $CD^2FA$ [30] and PDFA [31] with the aim of maintaining a minimum throughput of uncompressed or traditional DFA and reduction of memory space as well, are presented one after another.

In most of the researches for RE matching traditionally have been attempted to process regular expressions as a raw input have been tried to optimized the NFA or DFA structures. But in [23], an architecture has been proposed which is capable of detecting attack signatures (containing patterns and RE) without using complex RE-dependent NFA or DFA structures. The proposed architecture has fixed automata for analyzing the application protocols and other required features can be effectively compared in table structures.

However GPP-based architectures are more flexible and more reusable in comparison with other hardware architectures, but this kind of platform suffers from performance limitation. Design of efficient ASIP is a shortcut to overcome this limitation and use the benefits of flexibility and reusability of processors. In the next section, a framework is presented for analyzing and performance evaluation of a set of popular ASIPs for development of efficient Network based Intrusion Detection System.

## 3. PROPOSED FRAMEWORK

For performance evaluation of V850, OR1K, MIPS32 from MIPS series, ARM7TDMI from ARM series and PowerPC32 from PowerPC microprocessors a framework is required. Figure 2. illustrates the proposed framework and its work flow.

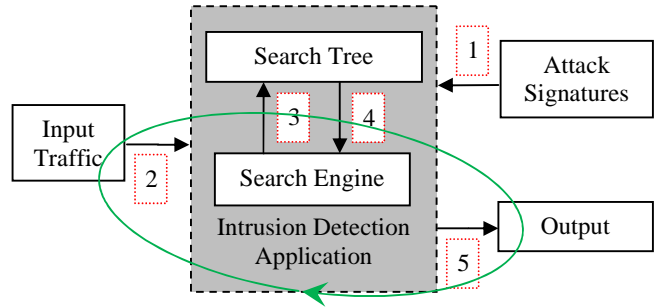

Figure 2. IDS proposed framework and execution phases

This framework consists four main modules: Attack signatures, Input traffic, Intrusion detection application, and output.

*Attack signatures module* – As mentioned in previous section, Snort is one of a software IDS that has a huge database of rules an attack signatures. This database is used in the attack signature module.

*Input traffic module* – Special network traffic is required for applying to the framework and extracting the results. By the same input traffic the results can be compared with each other.

*Intrusion detection application module* – This module has two parts. The first one is Search tree and the second one is Search engine. Based on Snort database, an Aho-Corasick search tree is created in this part. Aho-Corasick string matching algorithm is a string search algorithm (Important class of string algorithm which attempts to find the location of one or several strings that are named pattern in the longer string or text) which was invented by Alfred V. Aho and Margaret J. Corasick [32] in 1975. The search complexity of $T[1 \ldots m]$ with the Aho-Corasick automaton is $O(m + z)$ that $z$ is the number of occurred patterns in $T$. Because of linear search that increases the search speed, Aho-Corasick automaton is used in this research. In [33], all of the documents for data structure, files and functions of Aho-Corasick are available in summary. Search engine part has three duties. First is getting the input traffic and the second is applying this traffic to the Search tree and receiving the result of finding or not finding the attack pattern and the third is sending the result and some more information about the input traffic to Output module.

*Output module* – If there is a definite attack pattern in input packet payload, a report is recorded for the occurred attack in an output file.

## 4. OPTIMIZATION

A compiler is likely to perform many or all of the following operations: lexical analysis, preprocessing, parsing, semantic analysis (Syntax-directed translation), code generation, and code optimization.

In this research, code optimization as one of the compiler operations is used for increasing the performance.

Compilers bridge source programs in high-level languages with the underlying hardware. A compiler requires 1) determining the correctness of the syntax of programs, 2) generating correct and efficient object code, 3) run-time organization, and 4) formatting output according to assembler and/or linker conventions. A compiler consists of three main parts: the frontend, the middle-end, and the backend [34].

The front end checks whether the program is correctly written in terms of the programming language syntax and semantics. The middle end is where optimization takes place. The back end is responsible for translating the Intermediate Representation (IR) from the middle-end into assembly code.

This front-end/middle/back-end approach makes it possible to combine front ends for different languages with back ends for different CPUs. Practical examples of this approach are the GCC, LLVM, and the Amsterdam Compiler Kit, which have multiple front-ends, shared analysis and multiple back-ends.

### 4.1. GCC

The GCC is a compiler system produced by the GNU Project supporting various programming languages.

GCC has been ported to a wide variety of processor architectures, and is widely deployed as a tool in commercial, proprietary and closed source software development environments. GCC is also available for most embedded platforms, for example Symbian (called *gcce*), AMCC and Freescale Power Architecture-based chips [35].

GCC 1.0 which only handled the C programming language was released in 1987, and the compiler was extended to compile C++ in December of that year. Front ends were later developed for Fortran, Pascal, Objective-C, Java, and Ada, among others. The current stable version of GCC is 4.6.1, which was released on June 27, 2011.

### 4.2. Optimization with GCC Predefined Optimization Levels

Optimizations in GCC are done by the flags that use in gcc command line. –f<optimization name> is used for activating a flag and –fno–<optimization name> is used for deactivating a flag in command line. The GCC also has its own predefined levels of optimization [36] which begin with –O and include: –O or –O1, –O2, –O3, –O0 and –Os.

## 5. IMPLEMENTATION AND EXPERIMENTAL RESULTS

Open Virtual Platform (OVP) [37] is a fast simulation with open source and free resource model and has Application Program Interfaces (API). The focus of OVP is to accelerate the adoption of the new way to develop embedded software, especially for System-on-Chip (SoC) and Multiprocessor System-on-Chip (MPSoC) platforms. OVP uses libraries of processor and behavioral models, and APIs for building the own processors, peripherals and platforms. OVP is flexible and is free for noncommercial usages. This simulation is a product of 2008 and used in this research.

The implementation of each proposed framework module is explained below.

*Implementation of Attack signatures module* – Special format is used for writing the attack signatures in Snort. These signatures are divided to rule header and rule options sections logically [38]. Rule header includes rule operation, protocol, source and destination IP addresses and their netmask and source and destination port information. Rule option section includes alert messages and information that should be considered about some parts of packet. 15597 Snort rules are used in this research.

*Implementation of Input traffic module* – The Cyber Systems and Technology Group of MIT Lincoln Laboratory, under Defense Advanced Research Projects Agency (DARPA) and Air Force Research Laboratory sponsorship, has collected and distributed the first standard corpora for evaluation of computer network intrusion detection systems [39]. The 1998 DARPA evaluation was designed to find the strength and weaknesses of existing approaches and lead to large performance improvements and valid assessments of intrusion detection systems. This research uses five hundred thousand packets from simulation output traffic of the third week on Thursday, Lincoln Laboratory in 1999.

*Implementation of Intrusion detection application module* – This module was written in C programming language. In creating the search tree, according to attack rules in Snort that are divided to four sections consist of TCP, IP, ICMP and UDP based on their protocols, four search tree are made with the same names. Next, by calling the Create_Aho_Tree, reading the attack rules which already were downloaded from Snort.org site are started from *.rules files. In search engine, darpa_traffic function is called for inspecting the incoming traffic. This function reads the packets and then acquires the payload of them and determines their protocols, including ICMP, IP, TCP and UDP. By calling ahocorasick_KeywordTree_search _helper function, the packet payload is searched in corresponding tree. In this projects, only 6 (content, nocase, offset, depth, distance and within) of 37 rule options are examined.

*Implementation of Output module* – An output text file is created in this module that includes a report of attacks. This report includes source and destination IP addresses, source and destination ports, the packet payload and alert message of found attack signature in packet.

In this research, version 2/23/2011 of OVP simulator program is used on a laptop with Windows XP SP2, 1.60 GHz CPU and 512 MB RAM. The simulation has used the basic microprocessors without cache and pipeline. All microprocessors have the same nominal speed, and are equal to 100 MHz.

Execution of intrusion detection application on V850 is encountered message "Heap and stack collision" because of memory shortage so this microprocessor is ignored. Figure 3. illustrates the number of assembly instructions than the number of incoming packets for each microprocessor.

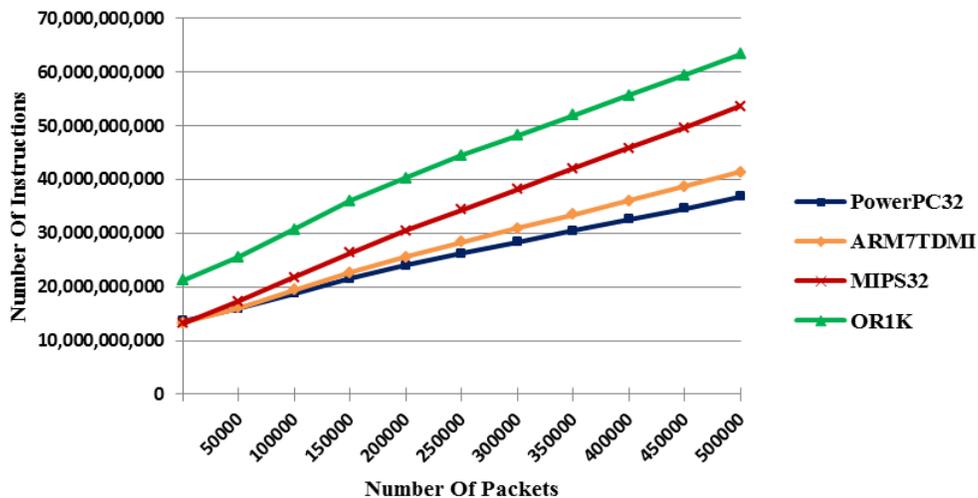

Figure 3. Execution graph of intrusion detection application on four microprocessors

When there aren't any incoming packets, the graph shows the number of assembly instructions that are needed for making attack signature trees. However in making search tree section, ARM7TDMI, MIPS32 and PowerPC32 have almost the same performance but gradually with arrival of packets to the system, PowerPC32 microprocessor will be better. Because all of microprocessors have the same speed (100 MHz) to execute the instructions so run-time of intrusion detection application for five hundred thousand packets is explained in Table I. Run-time of intrusion detection application in making search tree is not important because this tree is made just for one time, so Table 1. just shows run-time of intrusion detection application without the time that is needed for making search tree.

Table 1. Run-time of intrusion detection application for five hundred thousand packets.

| Microprocessors | Run-Time |
|---|---|
| PowerPC32 | 231.75 s |
| ARM7TDMI | 280.95 s |
| MIPS32 | 404.66 s |
| OR1K | 420.91 s |

The performance of microprocessors are checked again for intrusion detection application but this time the optimization levels –O1, –O2 and –O3 are used. These three optimization levels reduce the run-time of applications. The performance is evaluated relative to the –O0 level which is the level without optimization. –O0 level results were shown in Table 1.

Table 2. shows the performance percentage of microprocessors with optimization levels relative to –O0 level.

Table 2. Performance increase percent of microprocessors by using predefined optimization levels for five hundred thousand packets.

| Microprocessors | O1 to O0 | O2 to O0 | O3 to O0 |
|---|---|---|---|
| PowerPC32 | 20.32 % | 15.47 % | 15.34 % |
| ARM7TDMI | 11.91 % | 13.57 % | 13.41 % |
| MIPS32 | 8.23 % | 8.89 % | 8.78 % |
| OR1K | 28.36 % | 28.10 % | 27.87 % |

In order to increase performance, by focusing on ARM7TDMI microprocessor, it's attempted to increase the run-time of intrusion detection application on this microprocessor. For this purpose, compiler improvement is used like the previous part.

As mentioned in Table 2. O2 level has the best functionality in the second section of intrusion detection application (Searching attack signature in packet payloads) in ARM7TDMI. For improving the performance, O2 level is intended as a baseline and by adding some other flags to this level, a new combination is presented that works better than O2 level in ARM7TDMI.

As regards the intrusion detection application has too loops with many iteration and the long jumps, first the flags in gcc that optimize the mentioned issues, are collected. By frequent running the intrusion detection application in binary combination (O2 with another flag), triad combination (O2 with two other flags) and so on, the following combination is obtained:

```
-O2  -freduce-all-givs  -fmove-all-movables   -mcpu=arm7  -fnew-ra
-fno-expensive-optimizations   -fno-force-mem
-fno-guess-branch-probability   -fno-if-conversion2  -fno-crossjumping
```

Table 3. shows the performance improvement (O2 to O0 and Offered to O0) for execution of intrusion detection application in ARM7TDMI. According to Table 3, proposed combination works 4.53 percent better than O2 level.

Table 3. Performance increase percent of ARM7TDMI microprocessor in O2 and offered level.

| Optimization Level | Improvement percentage to O0 |
|---|---|
| O2 | 13.57 % |
| Offered | 18.10 % |

The use of compiler optimization levels will not always improve the performance. Using the C functions, specially the functions that involve to string such as strcmp, strlen, strlwr and strstr in the written IDS wasn't increased more than 0.5% by predefined optimization levels. but with eliminating mentioned functions (If it is possible) or rewriting them with loops and conditions or using equivalent functions but with better performance led to performance range is putted in 8.23% to 28.36%. For example in the second part of intrusion detection application namely inspecting the packet payloads, Boyer Moore algorithm was used instead of strstr that almost works 50% better than it.

Unlike existent documents that know optimizing in O3 level is more than O1 and O2, this research shows that optimization of these levels are not deterministic and depends on application.

## 6. CONCLUSION

In this paper different kinds of hardware architectures for network based intrusion detection systems have been reviewed. These architectures have been categorized based on hardware platforms and the structures of hardware designs. The presented taxonomy shows the lack of sufficient study on application specific instruction-set processor as a flexible, reusable, high performance and low cost solution for deployment of NIDS.

In addition, a framework is presented for analyzing and performance evaluation of pattern matching for intrusion detection on specific microprocessors including V850, OR1k, MIPS32, ARM7TDMI and PowerPC32.

For optimization of execution time on the specific processors, different level of optimization have been considered. This study shows that just by using the predefined optimization levels, the performance of mentioned microprocessors can be increased between 8.23% to 28.36 which is fairly substantial. This is important because optimization is performed with lower cost and easier than other solutions such as hardware design changing. For more minimization of run time of pattern matching application, a combination of compiler optimization flags is introduced for ARM7TDMI. Offered optimization level improved run time of NIDS application around 18.10%.

## REFERENCES

[1]  J. P. Anderson, (1980) "Computer security threat monitoring and surveillance", Technical report, James P. Anderson Company, Fort Washington, Pennsylvania, April.

[2]  D. Denning, (1987) "An intrusion-detection model", IEEE *Transactions on Software Engineering* 13 (2) 222-232.

[3]  Zebo Peng, Eslab and LiTH, (2007) "Application Specific Instruction Processor Architecture", http://en. scientificcommons.org/zebo_peng .

[4]  Jose M. Gonzalez , Vern Paxson , Nicholas Weaver, (2007) "Shunting: a hardware/software architecture for flexible, high-performance network intrusion prevention", Proceedings of the 14th ACM *conference on Computer and communications security*.

[5]  G. Vasiliadis, M. Polychronakis, and S. Ioannidis, (2011 ) "MIDeA: A Multi-Parallel Intrusion Detection Architecture," *In Proc. CCS,*.

[6]  M. Roesch, (1999) "Snort - lightweight intrusion detection for networks", In Proc. LISA99, the 13th Systems Administration Conference.

[7]  V. Paxson, (1998) "Bro: a system for detecting network intruders in real-time", In Proc. 7th USENIX Security Symp., San Antonio, TX.

[8]  R. S. Boyer and J. S. Moore, (1977) "A fast string searching algorithm", Communications of ACM, 20(10):761-772.

[9]  V. Aho and M. J. Corasick, (1975) "Efficient string matching: An aid to bibliographic search", Communications of ACM, 18(6):333-340.

[10] L. Tan and T. Sherwood, (2005) "A high throughput string matching architecture for intrusion detection and prevention", In ISCA'05:32nd Annual International Symposium on Computer Architecture, pp. 112-122.

[11] I. Sourdis, D. Pnevmatikatos, (2004) "Pre-decoded CAMs for efficient and high-speed pattern matching", Proc. FCCM.

[12] Z. K. Baker, V. K. Prasanna, (2004) "A Methodology for synthesis of efficient intrusion detection systems on FPGAs", Proc. FCCM.

[13] T. Katashita, Y. Yamaguchi, A. Maeda, and K. Toda, (2007) "FPGA based Intrusion Detection System for 10 Gigabit Ethernet", IEICE Transaction on Information and Systems, vol. 90, no. 12, pp. 1923-1931.

[14] L. Bu, J. A. Chandy, (2004) "FPGA based network intrusion detection using content addressable memories", Proc. FCCM.

[15] G. Vasiliadis, S. Antonatos, M. Polychronakis, E. P. Markatos and S. Ioannidis, (2008 ) "Gnort: High Performance Network Intrusion Detection Using Graphics Processors", In Proc. 11[th] Int. Symposium on Recent Advance in Intrusion Detection.

[16] N. Cascarano, P. Rolando, F. Risso, and R. Sisto, (2010) "iNFAnt: Nfa pattern matching on gpgpu devices", ACM SIGCOMM Comput. Commun. Rev., 40:20–26.

[17] S. Dharmapurikar, J. W. Lockwood, (2006) "Fast and Scalable Pattern Matching for Network Intrusion Detection Systems", IEEE J. On Selected Areas in communication, vol. 24, no. 10.

[18] M. Aldwairi, T. Conte, P. Franzon, (2005) "Configurable string matching hardware for speedup intrusion detection", In Proc. ACM SIGARCH Computer Architecture News, vol. 33, no. 1, pp. 99-107.

[19] E. Spitznagel, D. Taylor, J.Turner, (2003) "Packet Classification Using Extended TCAMs", Proc. IEEE ICNP.

[20]   K. Pagimatzis, A. Sheikholeslami, (2006) "Content-addressable memory (CAM) circuits and architectures: A turorial and survey", IEEE Journal of Solid-State Circuits, vol. 41, no. 3, pp. 712-727.

[21]   M. Gokhale, D. Dubois, A. Dubois, M. Boorman, S. Poole and Vic Hogsett, (2002) "Granidt: Towards Gigabit Rate Network Intrusion Detection", In Proc. International Conference on Field-Programmable Logic and its Applications (FPL).

[22]   F. Yu, R. Katz, T. V. Lakshman, (2004) "Gigabit Rate Packet Pattern-Matching Using TCAM", In Proc. IEEE ICNP, pp. 174-183.

[23]   M. A. Taherkhani, M. Abbaspour, (2009) "An Efficient Hardware Architecture for Deep Packet Inspection in Hybrid Intrusion Detection Systems", In Proc. $4^{th}$ Int. Conf. on Communications and Networking in China (ChinaCom09), August 26-28.

[24]   J. Moscola, J. Lockwood, R.P. Loui, and M. Pachos, (2003) "Implementation of a Content-Scanning Module for an Internet Firewall", In Proceedings of FCCM.

[25]   Hongbin Lu, K. Zheng, B. Liu, X. Zhang and Y. Liu, (2006) "A Memory-Efficient Parallel String Matching Architecture for High Speed Intrusion Detection", In IEEE *Journal on Selected Areas in Communications*, Vol. 24, No.10.

[26]   T. Song, D. Wang, (2011) "Another CDFA Based Multi-Pattern Matching Algorithm and Architecture for Packet Inspection", In Proc. of 20th Int. Conf. on Computer Communications and Networks (ICCCN).

[27]   S. Dharmapurikar, P. Krishnamurthy, T. Sproull, and J. Lockwood, (2003) "Implementation of a Deep Packet Inspection Ciruit using Parallel Bloom Filters in Reconfigurable Hardware", In Proceedings of HOTi.

[28]   S. Yusuf, W. Luk, (2005) "Biwise optimized CAM for Network Intrusion Detection Systems", IEEE FPL.

[29]   S. Kumar et al, (2006) "Algorithms to Accelerate Multiple Regular Expressions Matching for Deep Packet Inspection", in ACM SIGCOMM'06, Pisa, Italy.

[30]   S. Kumar, J. Turner, and J. Williams, (2006) "Advanced algorithms for fast and scalable deep packet inspection", in Proc. of ACM/IEEE *Symposium on Architecture for Networking and Sommunications Systems (ANCS'06). New York, NY, USA*, ACM Press, pp. 81–92.

[31]   J. Jiang, X. Wang, K. He, B. Liu, (2010) "Parallel Architecture for High Throughput DFA-Based Deep Packet Inspection", In Proc. of IEEE Int. *Conf. on Communications (ICC)*, pp. 23-27.

[32]   http://en.wikipedia.org/wiki/Aho-Corasick_algorithm

[33]   Doxygen, (2004) "FFPT Reference Manual 1.3", http://ffpf.sourceforge.net

[34]   Compiler from Wikipedia, http://en.wikipedia.org/wiki/Compiler

[35]   GNU Compiler Collection from Wikipedia, http://en.wikipedia.org/wiki/GNU_Compiler_Collection

[36]   Optimize Options - Using the GNU Compiler Collection(GCC), http://gcc.gnu.org/online docs/gcc-4.1.1/gcc/Optimize-options.html

[37]   OVP Simulation, http://www.ovpworld.org/aboutovp.php

[38]   Sourcefire, Inc, (2010) "SNORT® Users Manual 2.9.0", The Snort Project.

[39]   http://www.ll.mit.edu/mission/communications/ist/corpora/ideval/data/index.html

**Authors**

Majid Nezakatolhoseini has graduated in M.Sc. Computer Engineering with major of Computer Architecture from Science and Research branch of Islamic Azad University of Iran at 2011. He got his B.Sc. degree in Computer Engineering with major of Computer Hardware from Najaf Abad Azad University of Iran at 1999.

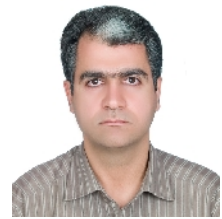

During his Master and Bachelor degrees, he also had researches on topics such as Network on Chip (NoC) and intelligent vehicle.
He has 7 years of experience in the field of Mobile communication networks in MCCI which is the biggest Mobile operator in Middle East. His specialty is in NOC (Network Operation Center).

Mohammad Amin Taherkhani received his BSc degree in Computer Engineering Hardware Engineering- from Amirkabir University of Technology (Tehran Polytechnic) in 2006. He also received his MSc degree in Computer Engineering - Computer Architecture- from Shahid Beheshti University (National University of Iran) in 2009. His research experiences and interests are: Hardware Accelerators for Network Security Applications, High Speed Intrusion Detection and Prevention Systems and Attack Plan Recognition.

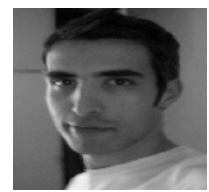